\documentclass[prl,twocolumn,showpacs,preprintnumbers,amsmath,amssymb]{revtex4-1}
\usepackage{dcolumn}

\usepackage[usenames,dvipsnames]{color}

\usepackage{graphicx}

\usepackage{amssymb}
\usepackage{amsmath}
\usepackage{bm}
\usepackage{epsfig}
\usepackage{ulem}

\newcommand {\be}{\begin{equation}}
\newcommand {\ee}{\end{equation}}
\newcommand {\bea}{\begin{eqnarray}}
\newcommand {\eea}{\end{eqnarray}}
\newcommand {\FIG}[1]{Fig. \ref{#1}}
\newcommand {\FIGS}[1]{Figs. \ref{#1}}
\newcommand {\EQ}[1]{Eq. (\ref{#1})}

\newcommand {\PRE}[1]{{Phys. Rev. E} {\bf {#1}}}

\newcommand {\PRL}[1]{{Phys. Rev. Lett.} {\bf {#1}}}

\begin{document}

\title{Phase transitions in Paradigm models}
\author{Huiseung Chae}
\author{Soon-Hyung Yook}
\author{Yup Kim} \email[Corresponding author:]{ykim@khu.ac.kr }
\affiliation{Department of Physics and Research Institute for Basic
Sciences, Kyung Hee University, Seoul 130-701, Korea}
\date{\today}

\begin{abstract}
In this letter we propose two general models for paradigm shift, deterministic propagation model (DM) and stochastic propagation model (SM).
By defining the order parameter $m$ based on the diversity of ideas, $\Delta$, we study when and how the transition occurs as a cost $C$ in DM
or an innovation probability $\alpha$ in SM increases. In addition, we also investigate how the propagation processes affect on the transition nature.
From the analytical calculations and numerical simulations $m$ is shown to satisfy the scaling relation $m=1-f(C/N)$ for DM with the number of agents $N$.
In contrast, $m$ in SM scales as $m=1-f(\alpha^a N)$.
\end{abstract}
\pacs{64.60.av, 89.65.-s, 87.23.Ge, 02.50.Le}
\maketitle

Transitions are ubiquitous in human history and in scientific activities as well as in physical systems.
Human history of civilizations has qualitatively distinguishable periods
from stone-age to contemporary civilizations, which depend on dominating themes such as philosophy, art, technology, etc.
In scientific activities such dominating themes correspond to disparate prevailing ideas or concepts such as chaos, complexity, nano, and string theory, etc., which are generally called as {\it paradigms}.
Tomas Kuhn said that the successive transition from one paradigm to another via revolution is the usual developmental pattern of mature science \cite{o1}.
This paradigm shift is also very similar to the adoption of a new discrete technology level.
Examples of such technological levels are operating system versions as Linux distributions and versions of recently-popular smart phones. 

To describe the appearance and disappearance of those paradigms, various models \cite{o4,o5,o6,o7,o9,o10} were suggested.
But, those models cannot describe the paradigm shift. 
Recently, an interesting model has been suggested by Bornholdt {\it et al.} to explain the dynamical properties of paradigm shifts \cite{Bornholdt}. In {\bf the Bornholdt model (BM)}, two essential mechanisms for the paradigm shift have been suggested. 
The first is the innovation process in which new ideas or paradigms are introduced.
The second is the propagation process in which idea of an agent possibly spreads to other agents. 
An important additional feature of BM is the memory effect that an 
agent never returns to the idea or the technological level once-experienced.
By the numerical study on a square lattice, Bornhodlt {\it et al.} have shown the existence of the ordered phase in which a
paradigm dominates for the small innovation probability $\alpha$ \cite{Bornholdt}.
In this ordered phase the pattern of the sudden emergence and slow decline of a new global paradigm repeats again and again.
However it is still an open fundamental question when and how this ordered phase disappears as $\alpha$ gets larger or approaches to 1. 

Furthermore the propagation of paradigms in BM is considered to occur locally and stochastically.
In contrast the propagation of ideas is generally successive and continuous or has the avalanches as can be seen from the spread of ideas through community networks, social network services and mass communication.
In addition, the propagation can occur deterministically originated from the differences (or gaps) of ideas (or technological levels) between the interacting pairs of agents \cite{Arenas1,Guar1,Ykim1}. 

To answer the raised questions and to investigate how the details of propagation processes affect the paradigm shift,
we provide two realistic and generalized models for paradigm shift, {\bf deterministic propagation model (DM)}
and {\bf stochastic propagation model (SM)}.
In DM the propagation is deterministically controlled by the difference of ideas, whereas the propagation is stochastically determined in SM.
Both models have avalanches of propagation.
By defining the order parameter, $m$, based on the diversity of ideas, $\Delta$, we analytically show that the disappearance of dominant paradigm can be mapped into the traditional order-disorder transition.
In DM we show that $m$ satisfies the
scaling relation $m=1-f(C/N)$, where $C$ is the propagation cost and $N$ is the total number of agents. 
In contrast, $m$ in SM follows the scaling relation $m=1-f(\alpha^a N)$, where $\alpha$ is the innovation probability. Here $f(x)$ is a scaling function satisfying $f(x)\sim x^b$ for $x \ll 1$ and $f(x)=1$ for $x \gg 1$. 
$m$ of BM is also proved to satisfy the same scaling relation as $m$ of SM. 
Thus, in DM the transition threshold $C^*$ scales as $C^*\simeq N$ and
 the transition probability in both SM and BM scales as $\alpha^* \sim N^{-1/a}$. The exponents $a$ and $b$ depend both on the models and on the underlying interaction topologies.
Therefore, from this work, we first provide the standard theoretical framework to understand  phase transitions and related phenomena in paradigm shift.

To be specific, let's assume that each agent resides on a node of a certain graph. 
At a given time $t$ each agent $i$ has a positive integer $r_i(t)$, which represents 
a particular idea or technological levels. 
Then at the time $t+1$, a randomly selected agent $i$ takes the innovation process with the probability $\alpha$ or propagates his idea to other agents with the probability $1-\alpha$.

In the innovation process at $t$, $r_i(t+1)$ of a randomly-chosen agent $i$ takes a discrete 
jump to be the smallest integer which has not been introduced to the whole system until the time $t$. 
To analyze phase transitions from the ordered phase to the disordered phase of the paradigm models,
we should first understand the model with $\alpha=1$, which we call {\bf the random innovation model (RIM)}\cite{remark}.
RIM cannot have the global paradigm and is always in the disordered phase. In RIM one can exactly calculate the diversity $\Delta(t)$,
which is defined as $\Delta (t) \equiv \sqrt{\left<r^2(t)\right>-\left<r(t)\right>^2}$, where $\left<r^k(t)\right> \equiv \left<[\sum_i r_i^k(t)]/N]\right>$ and $\left<...\right>$ means the average over realizations of models.
In RIM, a randomly selected agent $i$ at the time $t$ changes his idea $r_i(t) = t$. Let's denote $p \equiv 1/N$ and $q \equiv 1 - 1/N$, where $p$ is the selection probability of a particular agent.
Then the probability $P_t(r)$ that an arbitrary agent has the idea $r$ at $t$ is written as $P_t(r) = pq^{t-r}$ for $ 0 <r \le t$ and $P_t(0) = q^{t}$. 
In the limit $N \rightarrow \infty$, we get $\Delta(t) = N\sqrt{1-e^{-2t/N}-2(t/N)e^{-t/N}}$.
This result has been confirmed by simulation.
In the steady state (or $t \rightarrow \infty$),
$\Delta(t \rightarrow \infty) \equiv \Delta(\infty)=N$. 
$\Delta(\infty) =N$ corresponds to the disordered phase for $\alpha \rightarrow 1$ for paradigm models. Thus we take the order parameter $m$ for the phase transition of the paradigm models in the steady state as $m \equiv 1-\Delta(\infty) / N$. Then $m=0$ for the disordered phase 
and $m=1$ for completely ordered phase with $\Delta(\infty)=0$. 

We now consider two different paradigm models based on specifics of propagation process.
In DM the propagation is deterministically controlled by the cost in the following way.
A randomly selected agent $i$ propagates his idea $r_i(t)$ to
each nearest neighbor $j$ of $i$, i.e., $r_j(t+1) = r_i(t)$ at the time $t+1$ only if $r_i(t) -r_j(t) \ge C$.
Here $C$ is a constant which represents a cost or resistance to accept a new paradigm.
Then the propagation process triggers an avalanche; i.e., if
$r_j(t+1)$ is updated, then repeat the same propagation process for nearest neighbors of $j$. 
This propagation process is repeated until the inequality $|r_j(t)-r_i(t)| \le C $ is
satisfied for all the nearest pairs $\left<ij\right>$ in the system. 
\begin{figure}[ht]
\includegraphics[width=7cm]{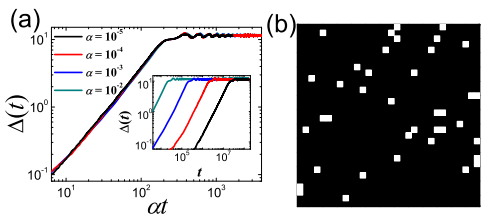}
\caption{(Color online) (a) Scaling plot of $\Delta(t)$ against $\alpha t$ of DM on a square lattice with $N=2^{12}$ and $C=82$. 
Inset: plot of $\Delta(t)$ against $t$.
(b) A snapshot of a steady state configuration of DM on the square lattice with the size $32 \times 32$. Black dots denote agents with a dominant idea $r_d$. White dots denotes those with ideas different from $r_d$.} \label{fig1}
\end{figure}
In DM, $\Delta(\infty)$ depends only on $C$ as shown in \FIG{fig1}(a), because $\alpha$ controls only the time $t_s$ taken for the system to arrive the steady state as $t_s \simeq 1/\alpha$. This result physically means that the system is in the steady state if the mean number of innovations, $\alpha t$, satisfies $\alpha t \gg C$ and the physical properties of the steady state depend only on $C$. 

\begin{figure}[ht]
\includegraphics[width=7cm]{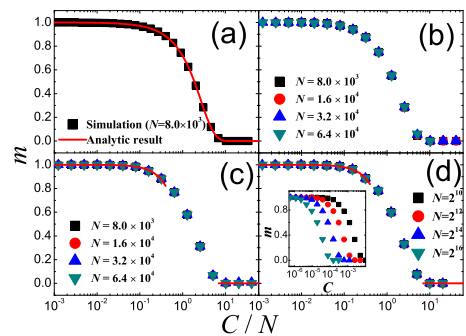}
\caption{(Color online) Scaling plots of $m$ against $ C/N$ of DM (a) on the complete graph with $N = 8.0 \times 10^3$, (b) on a scale-free network (c) on a random network and (d) on a square lattice. 
Curves in the figures show the analytic results (1) and (2).} \label{fig2}
\end{figure}

First we consider DM on the complete graph (CG).
Each agent on CG is a nearest neighbor of all the other agents.
Let's think a steady state configuration that ideas in the system spread in an integer interval $[r_{\min},r_{\max}]$ just after an innovation process at $t$. In the average sense $r_{\max}=\alpha t$. If $\alpha$ is small enough, propagation processes before the next innovation process drive the configuration into that with all $r_i > r_{\max} -C$, because of the propagation process initiated from an agent with $r_i=r_{\max}$. 
Then the probability $P_{r_{\max}}(r)$ that an agent has an idea $r$ in such configurations satisfies recursion relations $P_{r_{\max}+1}(r) = q P_{r_{\max}}(r)$ for $r_{\max}-C+1<r \le r_{\max}$ and $P_{r_{\max}+1}(r_{\max}+1) = p + qP_{r_{\max}}(r_{\max}-C+1)$.
From the recursion relations we obtain $P_{r_{\max}}(r) = pq^{r_{max}-r} + q^C P_{r_{\max}-C}(r-C) = ... = pq^{r_{\max}-r}/(1-q^C)$ and $\Delta(\infty) = N \sqrt{1-1/N- (C/N)^2 (q^{C})/(1-q^{C})^2}$ in the steady state.
In the large $N$ limit, $m$ thus satisfies 
\be \label{Eq1}
m = 1- g(C/N) ~ \left(g(x) = \sqrt{1-\left[\frac{x}{2} {\rm{cosech}}\left(\frac{x}{2}\right)\right]^2}\right).
\ee
Eq. (1) agrees very well with the simulation result as shown in \FIG{fig2}(a).
The ordered state of DM on CG has a peculiar physical property. Because $P_{r_{\max}}(r) \simeq (1-(r_{\max}-r)/N)/C \simeq 1/C$ for $C \ll N$, there doesn't exist a unique dominant idea, but $C$ ideas are nearly equally probable. This peculiar ordered state in the steady state comes from combination of the global connectivity of CG. 
In the sense that DM regards ideas with the idea difference $\delta r <C$ as the same idea, the ordered state on CG is physically plausible and understandable.

In contrast, there exists a unique dominating idea in DM on other graphs with local connectivities for $C \ll N$ as shown in \FIG{fig1}(b) and \FIG{fig2}. Thus we want to analytically show the existence of the ordered state with a dominating idea on the graphs.
Arbitrary nearest neighbor pair $\left<ij\right>$ of agents should satisfy the condition $|r_{i} - r_{j}| < C$ after a propagation process.
Let us think about the configuration with the $k$-th dominating macroscopic idea $r_d^{(k)}$. In the configuration the ideas in the system spread
in an integer interval $[r_{\min},r_{\max}]$ with $r_d^{(k)} \in [r_{\min},r_{\max}]$.
Now we want to show how the configuration with $(k+1)$-th dominating idea $r_d^{(k+1)}$ happens analytically.
As shown in \FIG{fig1}(b), the nodes (or sites) with $r_d^{(k)}$ form a macroscopic percolation cluster through the links (or bonds) of the graph and the nodes with $r \neq r_d^{(k)}$ form only isolated microscopic clusters. Thus the propagation process which changes the dominating idea happens the propagations only through the macroscopic percolation cluster. Therefore the configuration with the $r_d^{(k+1)}$ does not happen until the idea $r = r_d^{(k)} + C$ appears in the system.
After the idea $r = r_d^{(k)} + C$ appears, subsequent propagation processes through the macroscopic cluster make the configuration with $r_d^{(k+1)} (= r_d^{(k)} + C)$ appear before the next innovation process happens if $\alpha \ll 1/N$. The configurations with $r_d^{(k+1)} = r_d^{(k)} + C+1, ~ r_d^{(k)} + C+2$ and ... are also possible, but the probabilities that these exceptional configurations happen are at most order of $1/N^2$. So we neglect these exceptional configurations in the subsequent calculations. 
At the time of the paradigm shift the ideas in the system spread in the interval $[r_d^{(k+1)}-C+1, r_d^{(k+1)}]$. 
Then before the next paradigm shift, the configuration of the system can evolve to one
in which the ideas spread in the interval $[r_d^{(k+1)}-C+1,r_d^{(k+1)} + n_I]$. $n_I$ is the number of the innovations which happen before the $(k+2)$-th paradigm shift and $n_I <C$.
Generally the system in the steady state has a configuration with the ideas spread in the interval $[r_d-C+1,r_d + n_I]$. 

Now we consider the probability $P(r)$ that an agent has an idea $r$ in the steady state. Clearly $P(r)=0$ for $r \le r_d-C$ and $ r> r_d + n_I$. 
Furthermore, in the steady state $P(r)$ is expected to satisfy $P(r) \simeq 1/N$ for $r_d-C< r < r_d$ and $r_d < r \le r_d +n_I$, because an idea in the above intervals is originated from an innovation process. 
Thus we get $\Delta_{n_I}^2= (1/3N) ( C^3 + n_I^3)- (1/4N^2) (C^4 + n_I^4 - 2C^2 n_I^2)$.
From $\Delta^2(\infty) = C^{-1}\int _{0}^{C} \Delta_{n_I}^2 dn_I$, we get $\Delta(\infty)$ as $\Delta(\infty) = N \sqrt{(5/12)(C/N)^3 + (2/15)(C/N)^4}$. Therefore, $m$ for $C \ll N$ satisfies
$m = 1 - \sqrt{5/12} (C/N)^{3/2}$.
For $C \gg N$, DM reduces to RIM and $m=0$. Thus $m$ satisfies the scaling relation 
\be \label{Eq2}
m = 1 - f(C/N),
\ee
where $f(x) \sim x^{b}$ with $b=3/2$ for $x \ll 1$ and $f(x) = 1$ for $x \gg 1$. On CG we get the same scaling for $m$ with $b=1$.

To confirm the scaling relation on the graphs with local connectivity, we study DM by simulations on various graphs. The graphs used in this paper are a static scale-free network with the degree exponent $\gamma =2.5$ \cite{kahng}, and an Erd\"{o}s-R\'{e}nyi type random network, and a two-dimensional square lattice. 
To accord with the square lattice, the mean degree $\left<k\right>$ of the scale-free and random networks is set as $\left<k\right>=4$. The simulation datas of $m$ on each graph in \FIGS{fig2} are obtained by averaging over at least 1000 realizations. The scaling relation of $m$ with $b = 3/2$ or \EQ{Eq2} is confirmed by simulations on the random network and the square lattice as shown in \FIGS{fig2}(c) and (d).
In contrast, on a scale-free network with degree exponent $\gamma = 2.5$, we obtain the scaling relation with $b = 1.20(3)$ (\FIG{fig2}(b)), because the scale-free network has some aspect of global connectivity due to the hubs. 
Thus $C$ at which the phase transition occurs, $C^*$, scales as $C^* \simeq N$ on arbitrary graph. 

We now analyze SM in which the propagation process occurs stochastically.
In a propagation process of SM, a randomly selected agent $i$ always tries to propagate his idea $r_i(t)$ to all of nearest neighbors.
Explicitly, $r_j(t+1)$ of each $j$ of nearest neighbors to $i$ is made to be equal to $r_i(t)$ with the probability
$n_i/ N$, provided that $j$
never experienced $r_i(t)$ before. Here $n_i$ is the number of agents in the system which have the same idea as $r_i(t)$. In addition,
each agent $j$ whose idea is changed also propagates his changed idea to all of his nearest neighbors in the same manner
with the updated probability $n_i/N$, because $n_i$ increases as the propagation processes proceed.
This propagation process is repeated
until the propagation processes are terminated by the probability $(1-n_i/ N)$ or all the agents 
are tried to be propagated. Therefore, the propagation process of SM also has
the avalanche and an idea $r$ can spread
to the whole system at a given time.
Moreover, as we shall see, the scaling properties of SM on a graph with local connectivity are the same as those of BM \cite{Bornholdt}.

\begin{figure}[ht]
\includegraphics[width=7cm]{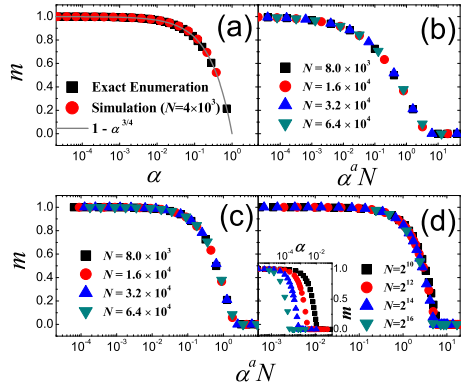}
\caption{(Color online) (a) Plot of $m$ against $\alpha$ of SM (a) on the complete graph.
The curve represents the analytic result $m=1-\alpha^{3/4}$. Scaling plots of $m$ of SM against $\alpha^a N$ on the scale-free network (b), on the random network (c) and on the square lattice (d).} 
\label{fig3}
\end{figure}

$m$ of SM on CG is analytically calculable, because an idea propagates to the whole system by single propagation process.
Let's consider a configuration that the ideas in the system spread in an integer interval $[r_{\min},r_{\max}]$
just before a propagation process. Then by the very next propagation process at the time $t_0$, an idea $r_d$ ($\in [r_{\min},r_{\max}]$) of a randomly selected agent in CG becomes the idea of all agents. Then until next propagation process, new ideas, $r_{\max}+1$, $r_{\max}+2$,... will appear by the subsequent innovation processes, because the system cannot have ideas experienced before.
The mean number of innovations after a propagation process until the time $t+t_0$ is
$\alpha t$. Let's think about a situation that all the propagation tries between $t_0$ and $t+t_0$ fail. Then at $t+t_0$, the probability $P_{t_0+t}(r)$ that an agent has the idea $r$ is written as $P_{t_0+t}(r) = 0$ for $r<r_d$, $P_{t_0+t}(r) = q^{\alpha t}$ for $r=r_d$ and $P_{t_0+t}(r) = pq^{\alpha t - (r-r_{\max})}$ for $r>r_{\max}$. Thus we can obtain $\Delta(t+t_0)$ easily. 
Since such propagation process happens again and again, $\Delta(\infty)$ is written as
$\Delta(\infty) = [\sum_{\delta r = 0}^{\infty} P(\delta r) \sum_{t=0}^{\infty}S(t)\Delta(t+t_0)]/[\sum_{t = 0}^{\infty} S(t)]$,
where $S(t)$ is the probability that no propagation processes happens until $t+t_0$ and $P(\delta r)$ of $\delta r =r_{\max}-r_d$ is the probability that a configuration with $r_i \in \{r_d, r_{\max}+1\}$ occurs at the very next innovation process after a propagation process.
Now we calculate $S(t)$. At $t+t_0$
the propagation probability is $(1-\alpha)(1-q^{\alpha t})/N$.
Then $S(t) = S(t-1) \left[1-(1-\alpha)(1-\exp(-\alpha t/N))/N \right]$ 
in the large $N$ limit. Thus we get
$S(t) = \exp\left[-(1-\alpha)t/N + (1-\alpha)(1-e^{-\alpha t/N})/\alpha \right]$.
$P(\delta r)$ can also be written as $P(\delta r) = (1-\alpha)p^2q^{\delta r} \sum_{t > \delta r/\alpha}^{\infty} S(t)$.
For $\alpha t \ll N$, $S(t) \simeq \exp(-t^2 \alpha(1-\alpha)/2N^2)$ 
and $\Delta(t+t_0) \sim N\sqrt{1/3}(\alpha t / N)^{3/2}$.
Therefore, $\Delta(\infty) \sim N \alpha^{3/4}$ for $\alpha t \ll N$ and
\be 
m(\alpha) = 1 - \alpha^{3/4}.
\ee
We also confirm Eq. (3) for arbitrary $\alpha$ 
by use of exact expressions of $\Delta(t+t_0)$, $S(t)$ and $P(\delta r)$ as shown in \FIG{fig3}(a). This result means that there always exists a dominating idea or the global paradigm on CG if $\alpha < 1$.
On the graphs only with local connectivity, the analytic approach as on CG to SM is hardly possible. Instead simulations are carried out. 
$\Delta(\infty)$ 
satisfies the scaling ansatz $\Delta(\infty)=h(\alpha^d N)$ very well.
As shown in \FIG{fig3}, $m$ satisfies the scaling function similar to that of DM as
\be
m = 1 - f(\alpha^a N), 
\ee
where $f(\alpha^a N)= h(\alpha^d N)/N$.
\{$a$, $b$\} are \{2.01(3), 0.49(2)\} on the scale-free network, \{1.15(2), 1.05(3)\} on the random network, \{1.10(2), 1.13(2)\} on the square lattice.
Thus the phase transition probability $\alpha^*$
scales as $\alpha^* \sim N^{-1/a}$ and $\alpha^*$ decreases as the global connectivity of graphs decreases.
Moreover the exponent $b$ increases as the global connectivity decreases. The scaling behavior of SM on the random network is nearly equal to that on the square lattice. This result means that the scaling behavior hardly depends on dimensionality of the graph, but depends on the connectivity. 

We also study $m$ of BM \cite{Bornholdt}. In BM, a randomly selected agent $i$ tries to propagate his idea to a randomly chosen nearest neighbor $j$ with the probability $n_i/ N$.
No further propagation processes are attempted in BM.
Since the propagation in BM is local, it is difficult to treat the model analytically even on CG. 
Thus  BM is studied numerically. From the simulations we confirm the same scaling behavior
$m = 1 - f(\alpha^a N)$ with $a = 1.10(2)$ and $b = 1.12(3)$ on any graph, especially on CG. The scaling behaviors are the same as those of SM on the square lattice. Since BM has only local propagation process on any graph and the propagation process does not use the connectivity of large scale or the global connectivity, even on CG, the scaling properties of BM are irrelevant to the dimensionality or the connectivity of the graph.
SM on the square lattice has also only local avalanches, and thus the scaling properties of SM on the square lattice are the same as those of BM. 
$\alpha^*$ of BM also scales as $\alpha^* \sim N^{-1/a}$ with $a \simeq 1.1$.

This work was supported by National Research Foundation of Korea (NRF) Grant funded by the Korean Government (MEST) (Grants No. 2011-0015257) and by Basic Science Research Program through the National Research Foundation of Korea(NRF) funded by the Ministry of Education, Science and Technology (No. 2012R1A1A2007430).

\end{document}